\documentclass[a4paper,10pt]{article}
\usepackage{amsmath,amssymb,graphicx}
\newcommand{\be}{\begin{equation}} % To start an equation
\newcommand{\ee}{\end{equation}} % To finish an equation
\newcommand{\bea}{\begin{eqnarray}} % To start an equation with many lines
\newcommand{\eea}{\end{eqnarray}} % To finish an equation with many lines
\title{Non extensivity of the QCD $p_{\mathrm{T}}$-spectra}
\author{T.~Bhattacharyya$^{1}$\footnote{bhattacharyya@theor.jinr.ru}~, %
J.~Cleymans$^{2}$\footnote{jean.cleymans@uct.ac.za}~, %
S.~Mogliacci$^{2}$\footnote{sylvain.mogliacci@uct.ac.za}~, \\ %
A.~S.~Parvan$^{1,3}$\footnote{parvan@theor.jinr.ru}~, %
A.~S.~Sorin$^{1}$\footnote{sorin@theor.jinr.ru}~ %
and O.~V.~Teryaev$^{1}$\footnote{teryaev@theor.jinr.ru}~}
\date{}
\begin{document}
\maketitle
\noindent $^1$ {\small Bogoliubov Laboratory of Theoretical Physics, Joint Institute for Nuclear Research, Dubna, Russia} \\ %
$^2$ {\small UCT-CERN Research Centre and Department of Physics, University of Cape Town, Rondebosch 7701, South Africa}\\ %
$^3$ {\small Department of Theoretical Physics, Horia Hulubei National Institute of Physics and Nuclear Engineering, Bucharest-M\u{a}gurele, Romania}
\vskip 0.1in
%%% ---------------------------------------------------------------------------------------------------- %%%
\noindent\rule{12cm}{0.4pt}
\begin{abstract} 
We establish a connection between the hadronic distributions, in proton-proton collisions at very high transverse momentum $p_{\mathrm{T}}$, obtained via perturbative QCD and the Tsallis non extensive statistics. Our motivation is that while the former is expected to be valid at extremely high momentum, due to asymptotic freedom, the latter has been very successful in describing experimental spectra over a wide range of momenta. Matching the non extensive statistics with the asymptotic $p_{\mathrm{T}}$ behaviour expected from QCD leads to the value of $q=1.25$.
\end{abstract}
\noindent\rule{12cm}{0.4pt}
%%% ---------------------------------------------------------------------------------------------------- %%%
\vskip 0.5in

The Tsallis distribution~\cite{tsallis} has been very successful in describing the particle spectra measured in high energy collisions~\cite{STAR, PHENIX1, ALICE_charged, ALICE_piplus, ALICE_PbPb, CMS1, ATLAS}. In absence of an equilibrium chemical potential $\mu$, at high enough energies, it is a distribution having only two parameters: The Tsallis $q$ parameter as well as the Tsallis temperature $T$. For relativistic classical systems, it is given by:
\bea
f = \left[1+(q-1) \frac{E}{T} \right]^{-\frac{q}{q-1}}
\label{tsallisdist}
\eea
The Tsallis distribution can be thought of as a superposition of the Boltzmann distributions, when the temperature in a system fluctuates following a gamma distribution~\cite{wilkprl}. The Tsallis parameter $q$ is then related to the relative variance of temperature, and controls the non additivity of the entropy relevant to the described system. In a number of situations, the non additivity implies some non extensivity even though the two concepts are strictly speaking different~\cite{HugoTouchette}. The Tsallis temperature $T$, which is the average of all the Boltzmann temperatures, is related to the Tsallis entropy $S_{\mathrm{T}}$ and the internal energy $U$ following~\cite{cleymansworkuthermcons}:
\bea
T^{-1}= \left.\frac{\partial S_{\mathrm{T}}} {\partial U}\right|_{V}
\eea
Hence, the form of the Tsallis distribution used in the Eq.~(\ref{tsallisdist}) is referred to as the thermodynamically consistent Tsallis distribution.

It should be noted that in the proton-proton collisions, this distribution has been able to describe particle production up to very high values of momenta~\cite{cleymansazmi}, and we seek for making a connection between the Tsallis non extensive statistics and perturbative QCD, as we expect the latter to be relevant in that $p_\mathrm{T}$ domain~\cite{Wong:2015mba,Wong:2013sca,Wong:2012zr}, by means of asymptotic freedom.

In order to have a QCD inspired description of the invariant cross section of the hadrons (focusing only on the mid rapidity region) over a wide range of transverse momenta $p_\mathrm{T}$, the following empirical formula had historically been introduced by Michael and Vanryckeghem~\cite{powerlaw1, powerlaw2} as well as by Hagedorn~\cite{hagedorn}:
\be
E \frac{d^3\sigma}{d^3p} = \mathcal{A} \left(1+\frac{p_\mathrm{T}}{p_0}\right)^{-c} %
\rightarrow \begin{cases} \mathrm{exp}\left(-c~\frac{p_\mathrm{T}}{p_0} \right)~~~\mathrm{for~} p_\mathrm{T}\ll p_0 \\ 
\left( \frac{p_\mathrm{T}}{p_0} \right)^{-c}~~~~~~~~~~\mathrm{for~} p_\mathrm{T}\gg p_0
\end{cases}
\label{TsallisQCD}
\ee
where $\mathcal{A}$ is a normalization factor, and $p_0$ acts as an effective temperature in the low $p_{\mathrm{T}}$ region while acting like a low momentum cut-off in the high $p_{\mathrm{T}}$ one.

We then notice that for dominant hard point-like scattering events, the value of the index $c$ is bounded as $c<4$~\cite{gunionprd,brodskyplb}. Moreover, we observe that the QCD cross section $ E \frac{d^3\sigma}{d^3p}$ is related to the invariant yield $E\frac{d^3N}{d^3p}$ of particles following:
\be
E \frac{d^3N}{d^3p} =\frac{E}{\sigma} \frac{d^3\sigma}{d^3p}~\Leftrightarrow~\frac{d^3N}{d^2p_{\mathrm{T}}dy} = %
\frac{1}{\sigma}\frac{d^3\sigma}{d^2p_{\mathrm{T}}dy}
\label{invyieldXsec}
\ee
By integrating the rightmost quantity of the above equation, or equivalently the right hand side of Eq.~(\ref{TsallisQCD}) over $p_{\mathrm{T}}$, we are left with a quantity related to experimentally measurable integrated yield $dN/dy$ which, like $\mathcal{A}$, plays the role of a normalization factor.

In first approximation, we can take the total cross section $\sigma$ to be independent of $p_\mathrm{T}$, and therefore the right hand side of the Eq.~(\ref{invyieldXsec}) will vary as $p_\mathrm{T}^{-4}$ at asymptotically high momentum. On the other hand, the leftmost quantity of Eq.~(\ref{invyieldXsec}), in the mid rapidity region and when parametrized by the Tsallis distribution, can be written as \cite{cleymansworkuthermcons}:
\be
E \frac{d^3N}{d^3p} = \frac{g V}{(2\pi)^3}~m_T~\left[1+(q-1) \frac{m_T}{T} \right]^{-\frac{q}{q-1}}
\label{TsallisParametrization}
\ee
where the transverse mass $m_T\equiv\sqrt{p_\mathrm{T}^2+m^2}$. % whose power goes as $p_\mathrm{T}^{-\frac{1}{q-1}}$ in the large $p_\mathrm{T}$ limit, provided we assume the Tsallis non extensive description to hold in that regime. 
If we take the asymptotically large limit $p_{\mathrm{T}}\rightarrow \infty$, then from Eqs.~(\ref{invyieldXsec}) and (\ref{TsallisParametrization}) we obtain:

\begin{eqnarray}
\lim_{p_{\mathrm{T}}\rightarrow \infty} \frac{E}{\sigma} \frac{d^3\sigma}{d^3p} = %
\frac{g V}{(2\pi)^3} p_{\mathrm{T}} \left[(q-1) \frac{p_{\mathrm{T}}}{T} \right]^{-\frac{q}{q-1}}
\label{tsallisasymp}
\end{eqnarray}
The large $p_\mathrm{T}$ part of the invariant yield, which is believed to have a `perturbative QCD origin' with hard scattering cross sections goes as $p_{\mathrm{T}}^{-4}$ . Also, to reiterate, we assume the total cross section $\sigma$ to be independent of transverse momentum in first approximation. Considering all these, we obtain the following equality valid for asymptotically high momentum:

\be
\lim_{p_{\mathrm{T}}\rightarrow \infty} \frac{E}{\sigma} \frac{d^3\sigma}{d^3p} \sim \frac{1}{p_{\mathrm{T}}^{4}}
\label{qcdasymp}
\ee
Matching the powers of $p_{\mathrm{T}}$ in the right hand sides of Eqs.~(\ref{tsallisasymp}) and (\ref{qcdasymp}), we get
\be
\frac{1}{q-1} = 4
\ee
which leads to
\be
q = 5/4 = 1.25
\ee
This above relation provides a natural upper bound to the Tsallis $q$ parameter, which is below the upper bound $q<4/3\approx 1.33$ required from thermodynamic consistency as pointed out in~\cite{BhattaCleMog}. We can also find out the Tsallis tempertaure corresponding to this upper bound with the help of Table 1 of Ref. \cite{ParvanEPJA} where the values of the fit parameters $q$ and $T$ for the pions are tabulated for energies ranging from $\sqrt{s}=$6.3 GeV (NA61/SHINE) to $\sqrt{s}=$7 TeV (CMS). The plot below shows the points in the $q-T$ plane and the function $q=1.572-0.005T$ which fits $q$ as a function of the central values of $T$. A relatively fair extrapolation leads to the value of temperature $T$ to be $\sim$ 59 MeV when $q=1.25$. A similar exercise may be done for other particles to find out the corresponding temperature values.

\begin{figure}[h]
\begin{center}
\includegraphics[width=3.2in]{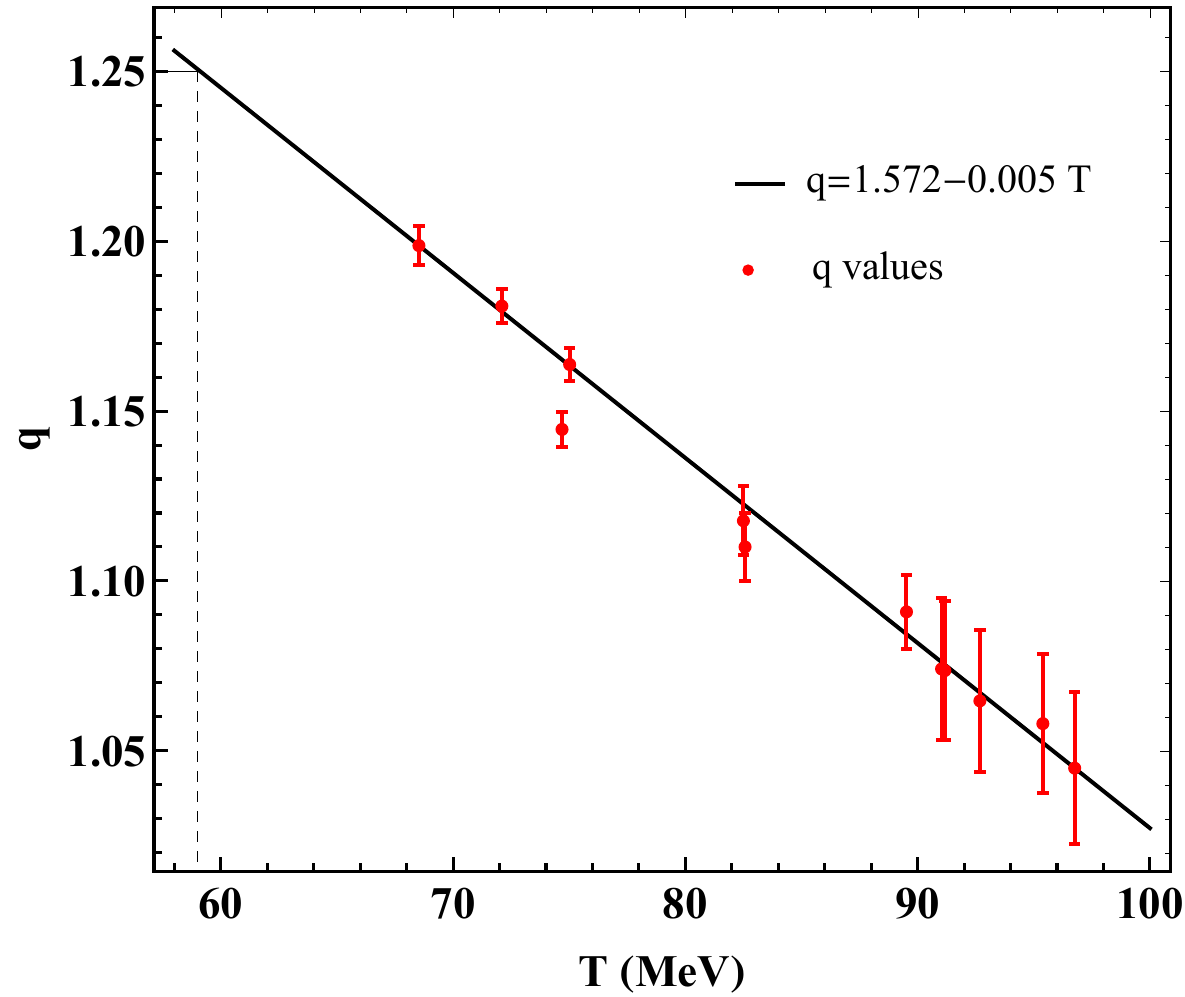}
\caption{$q$ vs $T$ for the $\pi^{-}$ particles. The vertical dashed line indicates $T=59$ MeV, when $q=1.25$.}
\label{nav1contourfig}
\end{center}
\end{figure}

Recently, a similar approach has been taken in Ref.~\cite{YalcinBeckSciRep} where the Tsallis statistics has been applied to the cosmic ray data to find out the $q$ value to be $q\sim 1.118$. But, the authors in this paper use a different version of the Tsallis distribution which at mid rapidity ($y=0$) and in the absence of chemical potential ($\mu=0$) is given by,
\bea
\left[1+(q-1)\frac{m_{\mathrm{T}}}{T}\right]^{-\frac{1}{q-1}}
\eea
The form of the distribution which we use in our manuscript has been shown, in Ref. \cite{cleymansworkuthermcons}, to be thermodynamically consistent.
Also, in the asymptotic limit, the power index of $p_T$ is taken to be 4.5, which in our case is 4 (according to Ref. \cite{Wong:2015mba} it may vary between 4 to 5). Hence, the approaches in the two papers albeit being similar, differ slightly in several important aspects.
%The remaining task is then to parametrize the total cross section $\sigma$ from the QCD side, before to be able to match the Tsallis parameter $T$ accordingly.
%
%In order to do so, we relate the rightmost quantity in Eq.~(\ref{invyieldXsec}) to the right hand side of Eq.~(\ref{TsallisParametrization}), leading to:
%\be
%\frac{1}{\sigma p_\mathrm{T}^4} = \frac{128 g}{\pi^3 p_\mathrm{T}^4}~T^5 V~~~\Rightarrow~~~\frac{1}{\sigma} = \frac{128 g T^5 V}{\pi^3}
%\ee

Finally, in order to conclude, we recall that we have established a connection between the hadronic distributions and the Tsallis distributions at very high transverse momentum domain where dominance of hard scattering has been assumed. From the dimensional analysis we obtain the upper bound of the Tsallis $q$ parameter to be $5/4=1.25$ which is below the one $q<4/3\approx 1.33$ proposed in \cite{BhattaCleMog} from the argument of convergence of the Tsallis thermodynamic quantities (like the number density for example). We however notice that while specifically fitting high $p_\mathrm{T}$ data (up to 20 GeV) from transverse momenta spectra at 7 TeV, such as that given by the ALICE collaboration~\cite{ALICE_7000}, we consistently obtain the same value for the $q$ parameter independent of the choice for the high $p_\mathrm{T}$ range lower bound. In addition, this value is also consistent with the one obtained in a recent investigation~\cite{bcmmp}, where the authors fitted the whole range including the very low $p_\mathrm{T}$ part. Although this is not a surprise, since we recall it has been shown that the Tsallis non extensive distribution can very well reproduce data over a wide range of momentum~\cite{cleymansazmi}, we can nevertheless translate this value of $q\approx 1.158$ to an index value of $c\approx 6.329$. This index value, which relies upon the thermodynamically consistent version of the Tsallis non extensive distribution suggests, unlike in~\cite{Wong:2015mba}, that there is no evidence for the dominance of the hard-
scattering processes. The transverse momentum distribution of jets in high energy $p-p$ and $\bar{p}-p$ experiments at high $p_{\mathrm{T}}$ and central rapidity shows a power-law behavior of $1/p_{\mathrm{T}}^n$ where $n\sim 4-5$. The power index for hadron spectra lies within the range of 6 to 10, slightly greater than those for jets, because hadrons are showering products from jets. Since hadronization is a non-perturbative process, we may argue that $n>4$ leads to non-perturbative effects. Hence, a deviation from the $p_{\mathrm{T}}^{-4}$ behaviour can be attributed to the higher order corrections as well as to the non perturbative effects and thus results in a modification in the upper bound of the $q$ value. Besides, a proper parametrization of the total cross section from the QCD side can enable one to establish a relationship between the parameters appearing in QCD and in the Tsallis statistics. We reserve this for a future work.

\vskip 0.4in
%%% ---------------------------------------------------------------------------------------------------- %%%
\renewcommand{\abstractname}{Acknowledgements}
\begin{abstract}
T.~B.~would like to acknowledge the financial support from the University Research Committee, University of Cape Town and the South African National 
Research Foundation. S.~M.~would like to acknowledge the financial support from the Claude Leon Foundation and the South African National 
Research Foundation. A.~S.~P. is supported by the funds of the joint research project of the JINR and IFIN-HH (protocol N 4543).
\end{abstract}

%%% ---------------------------------------------------------------------------------------------------- %%%

%%% ---------------------------------------------------------------------------------------------------- %%%

\begin{thebibliography}{100}
{\small%
\bibitem{tsallis} C.~Tsallis, J.~Stat.~Phys. {\bf 52} 479 (1988)
\bibitem{STAR} B.~I.~Abelev {\it et al.}, (STAR collaboration) Phys.~Rev.~C {\bf 75} 064901 (2007)
\bibitem{PHENIX1} A.~Adare {\it et al.}, (PHENIX collaboration) Phys.~Rev.~D {\bf 83} 052004 ( 2011)
\bibitem{ALICE_charged} K.~Aamodt {\it et al.}, (ALICE collaboration) Phys.~Lett.~B {\bf 693} 53 (2010)
\bibitem{ALICE_piplus} K.~Aamodt {\it et al.}, (ALICE collaboration) Eur.~Phys.~J~C {\bf 71} 1655 (2011)
\bibitem{ALICE_PbPb} B.~Abelev {\it et al.}, (ALICE collaboration) Phys.~Rev.~Lett. {\bf 109} 252301 (2012)
\bibitem{CMS1} V.~Khachatryan {\it et al.}, (CMS collaboration) JHEP {\bf 02} 041 (2010)
\bibitem{ATLAS} G.~Aad {\it et al.}, (ATLAS collaboration)  New~J.~Phys. {\bf 13} 053033 (2011)
\bibitem{wilkprl} G.~Wilk and Z.~W\l{}odarczyk, Phys.~Rev.~Lett. {\bf 84} 2770 (2000)
\bibitem{HugoTouchette} H.~Touchette, Physica~A {\bf 305} 84-88 (2002)
\bibitem{cleymansworkuthermcons} J.~Cleymans and D.~Worku, Eur.~Phys.~J.~A {\bf 48} 160 (2012)
\bibitem{Wong:2015mba} C.~Y.~Wong, G.~Wilk, L.~J.~L.~Cirto and C.~Tsallis, Phys.~Rev.~D {\bf 91} 114027 (2015)
\bibitem{Wong:2013sca} C.~Y.~Wong and G.~Wilk, Phys.~Rev.~D {\bf 87} 114007  (2013)
\bibitem{Wong:2012zr} C.~Y.~Wong and G.~Wilk, Acta Phys.~Polon.~B {\bf 43} 2047 (2012)
\bibitem{cleymansazmi} M.~D.~Azmi, J.~Cleymans, J.~Phys.~G {\bf 41} 065001 (2014)
\bibitem{powerlaw1} C.~Michael and L.~Vanryckeghem, J.~Phys.~G {\bf 3} L151 (1977)
\bibitem{powerlaw2} C.~Michael, Prog.~Part.~Nucl.~Phys.~{\bf 2} 1 (1979)
\bibitem{hagedorn} R.~Hagedorn, Riv.~Nuovo~Cim. {\bf 6 N10}, 1 (1983)
\bibitem{gunionprd} R.~Blankenbecler, S.~J.~Brodsky and J.~Gunion, Phys.~Rev.~D {\bf 12} 3469 (1975)
\bibitem{brodskyplb} S.~J.~Brodsky, H.~J.~Pirner and J.~Raufeisen, Phys.~Lett.~B {\bf 637} 58 (2006)
\bibitem{BhattaCleMog} T.~Bhattacharyya, J.~Cleymans and S.~Mogliacci, Phys.~Rev.~D {\bf 94} 094026 (2016)
\bibitem{ParvanEPJA} A.~S.~Parvan, O.~V.~Teryaev, J.~Cleymans, Eur. Phys. J. A  {\bf 53} 102 (2017)
\bibitem{YalcinBeckSciRep} G.~C.~Talcin and C.~Beck, Scientific Reports {\bf 8}, 1764 (2018)
\bibitem{ALICE_7000} J.~Adam {\it et al.} (ALICE collaboration), Eur.~Phys.~J.~C {\bf 75} 226 (2015)
\bibitem{bcmmp} T.~Bhattacharyya, J.~Cleymans, L.~Marques, S.~Mogliacci and M.~W.~Paradza, J. Phys. G {\bf 45} no.5, 055001 (2018)
}%
\end{thebibliography}
\end{document}